\documentstyle[12pt]{article}
\title{\large\bf An enhancement of neutral $tc$ transitions in
the model of dynamical breaking of the electroweak symmetry}
\author{\large\bf Arbuzov B.A.\footnote{E-mail: arbuzov@mx.ihep.su}
~and Osipov M.Yu.\\
{\it Institute for High Energy Physics, Protvino,}\\
{\it Moscow region, 142284, RF}}
\date{ }
\begin{document}
\newcommand{\bi}{\bibitem}
\newcommand{\be}{\begin{equation}}
\newcommand{\ee}{\end{equation}}
\newcommand{\beq}{\begin{eqnarray}}
\newcommand{\eeq}{\end{eqnarray}}
\maketitle
\bigskip
\begin{quote}
{ The problem of possible deviations from the Standard
Model is considered in the framework of a variant of
dynamical electroweak symmetry breaking.
It comes clear, that the parameters of the theory,
being obtained earlier and describing deviations from SM in
$Z\rightarrow \bar b b$ decay, are also consistent with the
existence of a nontrivial solution for vertex
$\bar t (Z,\gamma) c$. The occurrence of this solution leads to
a significant enhancement in neutral flavor changing transition
$t\rightarrow c$. The intensity of this transition is
connected with the $c$-quark mass, that leads to estimates of
probabilities of exotic decays $t\rightarrow c (Z,\gamma)$
(few \%) and of the cross-section of a single
$t$-quark production in process $e^+e^-\rightarrow t \bar c$
at LEP2 ($\simeq 0.03\,pb$ at $\sqrt{s} = 190\,GeV$).
The model is shown to be consistent with the totality of
the existing data; the predictions allow
its unambiguous check.}
\end{quote}
It is well-known, that the Standard Model (SM) of the electroweak
interaction agrees excellently with the totality of experimental
data with the only possible exception in decay
$Z\rightarrow b\,\bar b$~\cite{check}. However, variants of the
theory are considered, in which the electroweak symmetry breaking
is not due to the usual Higgs elementary scalars, but some
dynamical mechanism leads to the symmetry breaking.
In the present work
we consider the variant of the dynamical electroweak symmetry
breaking being proposed in~\cite{Arb1,ArbPav}, which is connected
with a selfconsistent mechanism of an appearance in the theory of
the additional gauge-invariant vertex of electroweak vector bosons'
interaction. This vertex effectively acts in the region of "small"
momenta, restricted by a cut-off $\Lambda$ being few $TeV$
by the order
of magnitude, which automatically appears in the theory. The vertex
of interaction of $W^+,\,W^-,\,W^0$ with momenta and indices
respectfully $p,\,\mu;\,q,\,\nu;\,k,\,\rho$ has the form
$$
\Gamma(W^+,W^-,W^0)_{\mu\nu\rho}(p, q, k)\,=\,
\frac{i \lambda g}{M_W^2}\,F(p^2,q^2,k^2)\,
\Gamma_{\mu\nu\rho}(p, q, k)\,;
$$
$$
\Gamma_{\mu\nu\rho}(p, q, k)\,=\,
g_{\mu\nu}(p_\rho (q k) -
q_\rho (p k)) + g_{\nu\rho}(q_\mu (p k) - k_\mu (p q))\,+
$$
\be
+\, g_{\rho\mu}(k_\nu (p q) - p_\nu (q k)) + k_\mu p_\nu q_\rho -
q_\mu k_\nu p_\rho\,.\label{vert}
\ee
$$
F(p^2,q^2,k^2)\,=\,\frac{\Lambda^6}{(\Lambda^2 - p^2)
(\Lambda^2 -q^2)(\Lambda^2 - k^2)}\,.
$$
Here $g$ is the gauge electroweak coupling constant,
$\lambda$ is the basic parameter of the model, nonzero value of which
follows from the solution of a set of equations for parameters of the
model~\cite{Arb1}. This solution leads to masses of gauge bosons $W$
and $Z$. We mean, that $W^0 = \cos\theta_W\,Z +\sin\theta_W\,A$ is
the neutral component of $W$ triplet. Note, that anomalous vertices
of the form~(\ref{vert}) are often considered in the framework of a
phenomenological
analysis of possible deviations from the SM~\cite{Hag1,Hag2}.
Equally with gauge bosons the $t$-quark has also a large mass.
The origin of its mass in our approach is connected with
anomalous vertex of its interaction with a photon
\be
\Gamma_\mu^{t}(p,q,k)\,=\,\frac{i e \kappa}{2 M_t}\,F(p^2,q^2,k^2)\,
\sigma_{\mu\nu}\,k_\nu\,;\label{magn}
\ee
As a result we obtain the theory, in which the initial symmetry is
broken, $W,\;Z,\;t$ acquire masses and all other quarks (and leptons)
are massless, the elementary Higgs scalars are absent and the main
distinction from SM consists in effective vertices ~(\ref{vert}),
~(\ref{magn}). These vertices lead, of course, to effects, which
differ the variant from the SM, that at the moment means the
existence
of limitations for parameters. Namely, the direct experiments (search
for $W$ pair production, $t$ pair production) give the following
experimental limitations~\cite{ArbSh,last}
\be
-0.31<\lambda<0.29;\qquad|\kappa|< 1.\label{restr}
\ee
On the other hand, vertices~(\ref{vert}),~(\ref{magn}) have to lead
to deviations from SM due to loop diagrams for other processes
especially
connected with the $t$-quark. Let us note, that the presence of
formfactors $F(p^2, q^2, k^2)$, $F_m(p^2,k^2)$ in the new vertices
results in the convergence of the loop integrals.
Just the problem of such deviations will be considered in the present
work. The significant effects appear also in weak interactions of
$b$-quarks. These effects were studied in the previous
work~\cite{ArbOs}, where the problem of a description of deviations
from the SM in parameters of decay
$Z\rightarrow\bar b b:\;R_b,\;A^b_{FB}$ was studied.
Namely, the measurement of these parameters give
$R_b = 0.2178\pm0.0011$ instead of SM value $0.2158$ and for
forward-backward asymmetry $A^b_{FB} = 0.0979\pm0.0023$ instead of
calculated $0.1022$~\cite{check}. The relative deviations
\beq
& &\Delta_b\,=\,\frac{R_b(exp) - R_b(th)}{R_b(th)}\,=
\,0.009\pm 0.005\,;\nonumber\\
& &\Delta_{FB}\,=\,\frac{A^b_{FB}(exp) - A^b_{FB}(th)}{A^b_{FB}(th)}
\,=\,-\,0.042\pm 0.023\,.\label{data1}
\eeq
indicate a possibility for a contradiction with SM predictions.
In the framework of our approach in studying of this problem
the effective vertex $\bar t W b$ is important. We have
assumed~\cite{ArbOs}, that here additional terms
of a magnetic dipole type are also present
\be
\Gamma_\rho^{tb}(p,q,k)\,=\,\frac{i g}{2 M_t}\,F_m(p^2,k^2)\,
\sigma_{\rho\omega}\,k_\omega\,\Bigl(\xi_+(1 + \gamma_5) +
\xi_-(1 - \gamma_5)\Bigr)\,;\label{tb}
\ee
where $p$, $q$ are respectfully momenta of $t$ and $b$ quarks,
$k$ is the momentum of $W$ and the formfactor looks like
$$
F_m(p^2, k^2)\,=\,\frac{\Lambda^4}{(\Lambda^2 - p^2)
(\Lambda^2 - k^2)}\,.
$$
It is important, that here enter both left-handed and right-handed
components of $b$-quarks.
Using the main vertices ~(\ref{vert}),~(\ref{magn}) we formulate
the set of equations for parameters of vertex~(\ref{tb}) in one-loop
approximation. For $\xi_+$ (left-handed $b$) we have an inhomogenous
equation, while for $\xi_-$ (right-handed $b$) we have a homogenous
one
\beq
& &\xi_+\,=\,-\,\frac{\lambda C \kappa}{24 \sqrt{2}\,F_1}\,
\biggl(1 - \frac{\sqrt{2}\,\kappa K }{8}\,\beta\,
\xi_+\biggr);\qquad
\xi_-\,=\,\frac{\lambda C K \kappa^2}{192\,F_2}\,
\beta\,\xi_-\,;\nonumber\\
& &F_1\,=\,1 - \lambda C\,\biggl(\frac{\kappa}{40} -
\frac{7}{48 \theta}\biggr)\,;\qquad
F_2\,=\,1 - \lambda C\,\biggl(\frac{\kappa}{40} -
\frac{1}{8 \theta}\biggr)\,;
\label{set}\\
& &\beta\,=\,
\frac{1}{4} + \frac{\lambda}{5 (1- \theta)};\qquad
C\,=\,\frac{\alpha \Lambda^2}{\pi M_W^2}\,;\qquad
K\,=\,\frac{\alpha \Lambda^2}{\pi M_t^2}\,.\nonumber
\eeq
Here and in what follows we use the abbreviated notation
$$
\theta\,=\,\sin^2\theta_W\,=\,0.23\,.
$$
It seems natural to decide, that the homogenuos equation for $\xi_-$
has trivial solution $\xi_- = 0$. However, with this solution it is
impossible to describe deviations~(\ref{data1}). It occurs, that
the satisfactory description can be achieved only if the equation for
$\xi_-$ has a nontrivial solution $\xi_- \not\equiv 0$.
This impose the following conditions on the parameters of the
problem
\be
\frac{\lambda C K \kappa^2}{192\,F_2}\,
\beta\,=\,1;\qquad \xi_+\,=\,-\,\sqrt{2}\,\kappa\,\theta.
\label{cond}
\ee
Emphasize, that condition~(\ref{cond}) by no means is the so called
fine tuning condition, because the parameters, which enter into it,
are just subjects for determination from the set~(\ref{set}).
Now parameter $\xi_-$ is ambiguous and we fix it and also the
parameter
\be
R\,=\,\frac{\xi_-}{\xi_+}\,;\label{R}
\ee
from the condition of correspondence with deviations~(\ref{data1}).
The simple calculations show~\cite{ArbOs}, that deviations~(\ref
{data1}) correspond to the following values of the parameters
\be
|R|\, =\, 2.45^{+0.25}_{-0.35}\,;\qquad
\xi_+\, = \,-\,0.043^{+0.013}_{-0.007}\,;\label{best}
\ee
Note, that value $\Lambda=4.5\,TeV$ of the effective cut-off
was used in the analysis, which corresponds to all the
selfconsistency conditionsof the model (see~\cite
{Arb1,ArbPav,ArbSh,ArbOs}) and the fine structure constant
$\alpha = 1/128$. With the existence of the nontrivial solution
the main parameters of the model acquire also fixed values
\be
\lambda\,=\,-\,0.23\pm 0.01\,;\qquad \kappa\,=\,0.13^{+0.02}_
{-0.04}\,;\label{lambda}
\ee
which are in the correspondence with experimental
limitations~(\ref{restr}).
Let us emphasize, that we can obtain the set of parameters
~(\ref{best}),~(\ref{lambda}), which gives the description of
data, including deviations~(\ref{data1}), only for the nontrivial
solution of set~(\ref{set}), which means an additional symmetry
breaking and leads to an appearance of terms, which are absent
in the initial theory in all orders of perturbation theory.
For example, the appearance of right-handed components of
the $b$-quark in vertex~(\ref{tb}) is impossible in the initial,
unbroken theory. Remind, that we consider the scheme, in which
all particles, but $W,\,Z,\,t$, are massless ($M_b = 0$ also).
In the present work we consider a possibility, that similar
phenomena can occur also for other vertices involving the $t$-quark.
Namely, we study vertex of interaction of neutral current $\bar t c$
and $W^0$. Such flavor changing current usually appears at the
two-loop level and gives small effects. However, nontrivial solutions
may lead here to an essential enhancement. Now, let us consider
vertex $\bar t W^0 c$
\be
\Gamma^{tc}_{\rho}(p,k)\,=\,\frac{i\, g}{2 M_t}\,F_m(p^2, k^2)\,
\sigma_{\rho \omega} k_\omega\,\Bigl(y_+ (1 + \gamma_5) +
y_- (1 - \gamma_5)\Bigr)\,;\label{tc}
\ee
where formfactor $F_m(p^2, k^2)$ is the same as in~(\ref{tb}),
$p,\,k$ are respectfully momenta of $t$ and $W^0$.
In the one-loop approximation vertex~(\ref{tc}) leads to an
appearance of new terms in vertex $\bar b W c\;(\Gamma^{bc})$,
which we schematically represent in the following form
\beq
& &\Gamma^{bc}\, = \,\biggl(\Gamma^{tb}\,\Gamma(W\,W\,W^0)\,
\Gamma^{tc}\biggr)\, +\, \biggl(\Gamma^{tb}_0\,\Gamma(W\,W\,W^0)\,
\Gamma^{tc}\biggr)\, + \nonumber\\
& &+\,\biggl(\Gamma^{tb}\,\Gamma_0(W\,W\,W^0)\,\Gamma^{tc}
\biggr)\,;\label{eqbc}
\eeq
where index "0" means the usual vertex of SM and we, of course,
mean the corresponding propagators between vertices and the
momentum integration $(d^4 q/(2 \pi)^4)$.
In expression~(\ref{eqbc}) there are terms with matrix structures
$\gamma_\rho$ and $\sigma_{\rho \mu}\,k_\mu$, denoting them
respectfully as $\Gamma_1^{bc}$ and $\Gamma_2^{bc}$, we write
down, again schematically, the following expression for
vertex~(\ref{tc})
\beq
& &\Gamma^{tc}\, =\, \biggl(\Gamma^{tb}\,\Gamma(W\,W\,W^0)\,
\Gamma^{bc}_0\biggr)\, +\,
\biggl(\Gamma^{tb}_0\,\Gamma(W\,W\,W^0)\,\Gamma^{bc}_2
\biggr)\, + \nonumber\\
& &+\,\biggl(\Gamma^{tb}\,\Gamma(W\,W\,W^0)\,\Gamma^{bc}_1
\biggr)\,;\label{eqtc}
\eeq
Performing the calculation of the loop integrals in expressions
~(\ref{eqbc}), (\ref{eqtc}), we obtain the following set of
equations for $y_\pm$
\beq
& &y_+\,=\,\biggl(-\,\frac{H^2}{8 \sqrt{2}}\,\biggl(\frac{\xi_+}{20}
- \frac{1}{12 \sqrt{2}}\biggr)\,-\,\frac{H S}{4}\,\xi_+^2 R^2\,+\,
\frac{5 H^2 \Lambda^2}{3072 M_t^2}\,\xi_+^2 R^2\biggr)\,y_+\,;
\nonumber\\
& &y_- = \biggl(-\,\frac{H S}{4} + \frac{5 H^2 \Lambda^2}{3072 M_t^2}
\biggr)\,\xi_+^2\,y_- + y_-^0\,;\; H = \frac{\lambda C}
{\theta}\,;\; S = \frac{K}{16 \theta}\,;
\label{settc}
\eeq
where
$$
y^0_-\,=\,-\,\frac{\lambda C \xi_+ U_{bc}}{8 \sqrt{2} \theta}\,;
\qquad |U_{bc}|\,=\,0.039\pm0.001\,;
$$
is the initial term for one-loop expression for vertex~(\ref{tc})
(ЇаЁ $y_\pm = 0$), and all other parameters are defined above.
Let us now look for a nontrivial solution of set~(\ref{settc}).
Provided $y_+\not\equiv 0$, the following condition is valid
\be
\frac{5 H^2 \Lambda^2}{3072 M_t^2}\,\xi_+^2 R^2\,
-\,\frac{H^2}{8 \sqrt{2}}\,\biggl(\frac{\xi_+}{20}
- \frac{1}{12 \sqrt{2}}\biggr)\,-\,\frac{H S}{4}\,\xi_+^2 R^2
\,=\,1\,.
\label{cond2}
\ee
It comes out, that condition~(\ref{cond2}) is satisfied by the same
values of the parameters as those obtained earlier. E.g., set of
the parameters
\be
\lambda = - 0.237;\qquad \xi_+ = - 0.039;\qquad R = 2.45\,;
\label{best1}
\ee
satisfies equation~(\ref{cond2}) and evidently is situated in the
range of accuracy of parameters~(\ref{best}), (\ref{lambda}). We
shall assume, that both conditions are fulfilled: (\ref{cond}) and
(\ref{cond2}). Their simultaneous fulfillment must not cause a
surprise. Indeed, condition~(\ref{cond}) is the equation for
two parameters: $\lambda,\,\xi_+$, and condition~(\ref{cond2})
is the equation for $\lambda,\,\xi_+,\,R$. Therefore, for the
fixed value of $R$ set of equations~(\ref{cond}) and (\ref{cond2})
gives two equations for two variables. Values~(\ref{best1}) for
parameters $\lambda$ Ё $\xi_+$ are just the solution of the set
for $R = 2.45$. Note once more, that this
solution describes data~(\ref{data1}).
For $y_-$ we obtain
\be
y_-\,=\,y^0_-\,\biggl(1 - \frac{1}{R^2} + \frac{H^2}
{8\sqrt{2}R^2}\Bigl(\frac{1}
{12\sqrt{2}}-\frac{\xi_+}{20}\Bigr)\biggr)^{-1}\,.\label{ymin}
\ee
The denominator in expression~(\ref{ymin}) is far from zero,
therefore the term with a normal chirality of the $c$-quark
in vertex~(\ref{tc}) turns to be small
due to a smallness of $y^0_-$. Substituting the parameters
being obtained we get estimate $|y_-| = 0.0012$. We shall see
below, that this estimate is, at least, two orders of magnitude
smaller, than a possible value of $y_+$, and so in what follows
we will not take into account contributions of $y_-$.
First of all, we can obtain upper limitation for the value of
parameter $y_+$ from data on decay $W^+\rightarrow \bar b c$.
$\Gamma^{bc}$ contains the contribution, corresponding to
equation~(\ref{eqbc}), that leads to a change in the probability
of this decay. Let us write down vertex $\Gamma^{bc}_\rho$,
taking into account this contribution
\beq
& &\Gamma^{bc}_\rho\,=\,\frac{g}{2\sqrt{2}}\,\Bigl( a\gamma_\rho +
b\gamma_\rho \gamma_5 + i\,c\sigma_{\rho\mu} k_\mu + i\,d
\sigma_{\rho\mu} k_\mu \gamma_5 \Bigr)\,;\nonumber\\
& &a = U_{bc}\,(1 + a_1)\,;\; b = U_{bc}\,(1 - a_1)\,;\;
c = d = -\,\frac{\lambda C y_+}{12 M_t \theta}\,;\label{Wbc}\\
& &a_1 = \frac{K\sqrt{2}\,\xi_+ y_+ R}{8 \theta}\,
\biggl(1 +\frac{5 \lambda}{12}\biggr)\,.\nonumber
\eeq
The modified decay width looks like
\be
\Gamma^M\,=\,\Gamma^{SM}\,(1 + \delta)\,;\qquad
\delta\,=\,\frac{2 a_1^2 +M_W^2 c^2}{2}\,;\label{widbc}
\ee
where $\Gamma^{SM}$ is the SM decay width.
The change of the width due to new terms has to
go into the accuracy of definition of the hadron width of
$W$. Taking into account the values of our parameters, we
obtain the following limitation
\be
|y_+|\,\leq\,0.9\,.\label{rest}
\ee
As we see from~(\ref{rest}), this limitation is not very
restrictive. It is interesting to obtain not the limitation,
but just the estimate of a possible value $y_+$. To achieve
this goal we use the contribution of vertex~(\ref{tc})
to the $c$-quark mass.
The initial $c$-quark mass in our approach is zero. Provided
the vertex of the type~(\ref{tc}) does not appear, the mass
remains to be zero. However, if the nontrivial
solution exists, there appear nonzero contributions to the
mass. The simplest terms correspond to two-loop diagrams,
which is described by the chains:
$c\rightarrow b(W^+)\rightarrow t(W^-)\rightarrow c(W^0)$ and
$c\rightarrow t(W^0) \rightarrow b(W^+) \rightarrow c (W^-)$.
Here in the brackets after transitions the sort of $W$ is
indicated and three gauge bosons are tied together by
vertex~(\ref{vert}). Transitions $c \leftrightarrow b$,
$b \leftrightarrow t$ are described by SM expressions,
whereas transitions $t \leftrightarrow c$ correspond to
vertex~(\ref{tc}). As a result we obtain the following
expression for $c$-quark mass in terms of a two-loop
integral
\beq
& &M_c\,=\,\frac{\alpha^2\,\lambda\,y_+\,U_{cb}\,\Lambda^4}
{32 \pi^2\,\sin^4\theta_W\,M_W^2\,M_t}\cdot I\,;
\label{cmass}\\
& &I = \frac{1}{\pi^4}\,\int\int\frac{p^2 (p^2 q^2 - (p q)^2)\,
d^4 p\,d^4 q}{(p^2)^2 (q^2)^2 (p-q)^2 ((p-q)^2 + 1) (p^2+1)^3
(q^2+1)} = \nonumber\\
& &= 0.48\,.\nonumber
\eeq
where substitutions $p\rightarrow p\Lambda,\,q\rightarrow q
\Lambda$ are made and the Wick rotation is performed, so that
the two-loop integral is calculated in the Eucleadean space.
The integral in~(\ref{cmass}) converges due to formfactors
(terms $(p^2+1)$ etc. in the denominator) and is equal
the produced number. The accuracy of relation~(\ref{cmass})
in our model is defined mainly by two points. Firstly, we
take into account the term with the maximal divergence
(without formfactors) of the forth order, neglecting
terms containing factor $M_t^2/\Lambda^2$. Secondly,
we could take into account also contribution of the additional
term~(\ref{tb}), which also leads to the fourth order
divergence. However, the factor
$\xi_+\,R \simeq 0.1$ is present in this case, so that
omitting of the term leads to additional uncertainty
about 10\%.
Relation~(\ref{cmass}) connects $y_+$ with $M_c$, all other
parameters here are either well-known or defined above.
Taking into account uncertainties in $M_c$ (which is the largest
one), in $M_t$ and in $U_{cb}$ and the above remark, we obtain
from~(\ref{cmass}) the possible interval for value $y_+$
\be
y_+\,=\,0.26\pm0.06\,.\label{yplus}
\ee
Now we can use value~(\ref{yplus}) for estimations of the
effects. Let us start with probabilities of exotic decays
of the $t$-quark: $t\rightarrow c\,Z,\; t\rightarrow c\,\gamma$.
Vertex~(\ref{tc}) immediately leads to the following expressions
for the decay widths
\beq
& &\Gamma(t\rightarrow c \gamma)\, = \,\frac{1}{4}\,\alpha\,M_t\,
y_+^2\,;\nonumber\\
& &\Gamma(t\rightarrow c Z)\, = \,\frac{g^2 (M_t^2-M_Z^2)}
{32\,\pi M_t^3\,(1- \theta)}
\,y_+^2\,\biggl(2 M_t^2- M_Z^2 - \frac{M_Z^4}{M_t^2}\biggr).
\label{tcZ}
\eeq
For value~(\ref{yplus}) the ratios of these widths to the total
one read
\beq
& &B_{\gamma} = BR\,(t\rightarrow c \gamma) = 0.0155\pm0.0055\,;
\nonumber\\
& &B_Z = BR\,(t\rightarrow c Z) = 0.053\pm0.023\,.\label{tcwid}
\eeq
Experimental limitations $B_{\gamma}\leq 3.2\%;\;B_Z\leq 33\%$
~\cite{exp} do not contradict to estimates~(\ref{tcwid}).
Furthermore, the estimates~(\ref{tcwid}) are not so far from
experimental limitations and thus have good prospects for a
check in a nearby future.
The next important process, in which vertex~(\ref{tc}) can
manifest itself, is the single $t$ ($\bar t$)-quark production
in $e^+e^-$ collisions above the threshold of $t\,\bar c$
production $\sqrt{s_{th}}\simeq 177\,ѓн‚$. Provided the energy
exceeds the threshold by few $GeV$ one may neglect $c$-quark
mass in the expression for the cross-section of process
$e^+\,e^-\rightarrow t\,\bar c$, which reads as follows
\beq
& &\sigma = \frac{\pi \alpha^2 y_+^2 (s - M_t^2)^2(s + 2 M_t^2)}
{2 M_t^2 s^2}\times\nonumber\\
& &\times\biggl(\frac{1}{s} +
\frac{1-4\,\theta}{4\,\theta (s - M_Z^2)}\, + \,
\frac{s (2 - 8\,\theta\, +\,16\,\theta^2)}{16\,\theta^2
(s - M_Z^2)^2}\biggr)\,.\label{cross}
\eeq
Process $e^+\,e^-\rightarrow\bar t\,c$ has the same cross-section,
so to estimate the cross-section for single $t\,(\bar t)$ production
we have to redouble expression~(\ref{cross}). Using values
$M_t = 176\,GeV$ and~(\ref{yplus}) we obtain the following estimates
for the cross-section of the single production at three energies
$(184\,Gev,\,192\,GeV,\,200\,GeV)$
\beq
& &\sigma(\sqrt{s}=184\,GeV)\,=\,\Bigl(0.012\pm 0.05
\Bigr)\, pb\,;\nonumber\\
& &\sigma(\sqrt{s}=192\,GeV)\,=\,\Bigl(0.037\pm 0.016
\Bigr)\,pb\,;\label{crossee}\\
& &\sigma(\sqrt{s}=200\,GeV)\,=\,\Bigl(0.066\pm 0.030
\Bigr)\,pb\,.\nonumber
\eeq
We see from the estimates, that for integral luminosity about
$200\,pb^{-1}$ one can expect few events of the single $t$-quark
production even at the energy $192\,GeV$. Therefore it seems,
that the study of a validity of the variant under consideration
is quite achievable for forthcoming experiments at LEP2 (see also
\cite{Obraz}).
One more observable effect, which we can estimate, is decay
$Z\rightarrow\bar c\,c$. One has to take into account four
one-loop diagrams with vertices~(\ref{tc}), (\ref{Wbc}).
With their account vertex $\bar c Z c$ looks as follows
\beq
& &\Gamma_\rho = \frac{g}{2 \cos\theta_W}\,\Bigl(a \gamma_\rho\,+\,
b \gamma_\rho \gamma_5\Bigr);\;a\,=\frac{1}{2} -
\frac{4}{3} \theta +b_1\,;\; b\,=\,\frac{1}{2}\,-\,b_1\,;
\nonumber\\
& &b_1 = \frac{K y_+^2}{4}\,\Biggl(\frac{1}{4 \theta} - \frac{1}{3}
+ D^2 \biggl(\frac{1}{2} - \frac{3}{4 \theta} +
\frac{3 (1- \theta)}{2 \theta} +
\frac{25 \lambda}{72 \theta}\biggr)\Biggr);\label{Zcc}\\
& &D\,=\,-\,\frac{\lambda\,C}{12 \sqrt{2} \theta}\,.\nonumber
\eeq
Vertex~(\ref{Zcc}) leads to the decay probability, which differs
from the SM prediction. We represent its ratio $R_c$ to the full
hadron width in the form $R_c = R_c^{SM} (1 + \Delta_c),
\,R_c^{SM} = 0.172$ and we have
\be
\Delta_c\,=\,\frac{36\,b_1^2\,-\,48 \theta\,b_1}
{9\,-\,24 \theta\,+\,32 \theta^2}\,.\label{dcc}
\ee
Expression~(\ref{dcc}) gives us value
\be
R_c\,=\,0.161\,\pm\,0.005\,;\label{rc}
\ee
which is to be compared with experimental number $0.1715 \pm 0.0056$.
The obtained value~(\ref{rc}) agrees with the experimental one,
however, it gives a marked difference with the SM. We could note
a remarkable point, that result~(\ref{rc}) agrees also with
experimental data, which have been advertised one and a half years
ago (see e.g. the discussion in review~\cite{check})
$R_c(old) = 0.1540 \pm 0.0074$ that has meant a considerable
deviation from the SM.
It is advisable to estimate the influence of changes in vertex
$\bar b c W$  for probabilities of $b$-quarks decays. In view of
this we calculate the width of decay $b\rightarrow c \bar \nu_e e$
\be
\Gamma(b\rightarrow c \bar \nu e)\,=\,
\frac{G^2 M_b^5 U_{bc}^2}{192 \pi^3}\biggl(1 - \frac{\lambda C y_+
M_b}{12 \theta U_{bc} M_t} + \frac{\lambda^2 C^2 y_+^2 M_b^2}
{360 \theta^2 U_{bc}^2 M_t^2}\biggr)\,.
\label{bce}
\ee
The expression in the brackets, which characterizes a deviation
from the SM, after substitution of the parameters gives,
depending on sign of $U_{bc}$ either $0.94$, or $1.10$.
Such deviations practically go into errors of definition of
$U_{bc}$. Thus the account of expression~(\ref{bce}) leads to
an insignificant shift of this parameter.
To conclude we mark few important points.
We have shown, that the model under consideration
describes quite satisfactorily the totality of experimental
data on the precise test of the electroweak theory,
including possible deviations from the SM in $Z$-decays.
The model also gives definite predictions, which can be looked
for; we draw attention to the following items.
1. Prediction for the constant of the anomalous triple vector
boson interaction $\lambda \simeq - 0.23$ can be tested at
LEP2~\cite{ee}.
2. Prediction for the cross-section of the single $t$-quark
production $\sigma \simeq 0.03\,pb$ at the energy of
$e^+ e^-$ collisions around $190\,GeV$ also can be studied
in the forthcoming experiments at LEP2. Exotic $t$-quark
decays at the level of accuracy $\simeq\,1\%$ are also
very important.
3. An improvement of the accuracy of data on decays
$Z \rightarrow \bar b b$, $Z \rightarrow \bar c c$
are also of a great interest. Maybe the second decay
is the most crucial for the test of the scheme.
The origin of masses of quarks (and leptons) is one of
the most important problems for the model. Here we have
shown how the $c$-quark mass can appear. The similar
mechanism can give also $b$-quark mass. In the present
work we do not consider this point, because presumably
one needs more high level of accuracy of calculations.
In fact, there are several contributions to the $b$ mass
of the same order of magnitude, and cancellation of large
numbers occurs. The uncertainties in parameters and
the approximations being used can decisively change a
result, in contrast to the case of the $c$-quark, where
we succeed in finding the leading term.
It is also very important for understanding of the structure
of the theory to study scalar excitements in systems
$W\,W$, $\bar t t$ etc., which could be interpreted as
composite Higgs particles. These problems will be considered
in the subsequent works.
The authors express gratitude to V.F. Obraztsov and S.R.
Slabospitsky for valuable discussions.

\end{document}